\documentclass[journal]{IEEEtran}
\usepackage{amsmath,amsfonts}
\usepackage{algorithm2e}
\usepackage{array}
\usepackage[caption=false,font=normalsize,labelfont=sf,textfont=sf]{subfig}
\usepackage{textcomp}
\usepackage{stfloats}
\usepackage{url}
\usepackage{verbatim}
\usepackage{graphicx}
\usepackage{cite}

\usepackage{makecell}

\usepackage{graphicx} 
\usepackage{amsmath}
\usepackage{amsmath}
\usepackage{amsmath, amssymb}
\usepackage{xcolor}	

\usepackage{booktabs}

\usepackage{algpseudocode}
\usepackage{graphics}
\usepackage{amsmath, amsthm}

\usepackage[caption=false,font=normalsize,labelfont=sf,textfont=sf]{subfig}

\newcounter{assumption}

\newcounter{theorm} 

\begin{document}
	
	\title{AI-Empowered Resource Allocation for Wirelessly Powered Pinching-Antenna Systems
	}

	\author{IEEE Publication Technology,~\IEEEmembership{Staff,~IEEE,}
	}

	\author{Saeid Pakravan, Mohsen Ahmadzadeh, Ming Zeng, Xingwang Li, and Fang Fang

\thanks{Copyright (c) 2026 IEEE. Personal use of this material is permitted. However, permission to use this material for any other purposes must be obtained from the IEEE by sending a request to pubs-permissions@ieee.org.}

\thanks{This work was supported in part by NSERC under Grants RGPIN-2021- 02636 and CRC-2022-00115, and in part by FRQNT under Grant 341270. (Corresponding author: Ming Zeng.)}

\thanks{S. Pakravan and M. Zeng are with the Department of Electrical and Computer Engineering, Laval University, Quebec City, QC, CA. email: saeid.pakravan.1@ulaval.ca;  ming.zeng@gel.ulaval.ca.}

\thanks{M. Ahmadzadeh is an Independent Researcher, Mashhad, Iran. email: M.ahmadzadehbolghan@gmail.com.}

\thanks{X. Li is with the School of Physics and Electronic Information
Engineering, Henan Polytechnic University, Jiaozuo 454000, China. email:
lixingwang@hpu.edu.cn.}

\thanks{F. Fang is with the Department of Electrical and Computer Engineering, Western University, London, Ontario, CA. email: fang.fang@uwo.ca.}

}
	\maketitle

	\begin{abstract}
		
This paper considers a multi-user system, where the users first harvest energy from the base station and then use the harvested energy to transmit information via non-orthogonal
multiple access (NOMA). A pinching-antenna array is adopted to assist the energy transfer and information transmission, owing to its ability to adapt to dynamic propagation
conditions. To enhance the system’s energy efficiency (EE), we formulate a joint optimization problem involving antenna positioning, transmit power control, and time-switching ratio selection. The problem is non-convex due to the coupled variables, nonlinear energy-harvesting characteristics, and uncertainties in user locations and battery states. To effectively solve this problem, a deep reinforcement learning based algorithm is proposed to autonomously learn near-optimal resource allocation policies in dynamic environments. Simulation results demonstrate that the proposed PA-assisted scheme achieves significant gains in EE compared with conventional fixed-antenna schemes.
		
	\end{abstract}
	
	
	\section{Introduction}

The exponential growth of wireless devices and the escalating demand for high data-rate, low-latency, and energy-efficient communications have imposed unprecedented challenges on modern wireless networks. Future beyond fifth-generation (B5G) and sixth-generation (6G) systems are expected to accommodate massive connectivity while ensuring high spectral efficiency and sustainable energy consumption, particularly in large-scale internet of things (IoT) deployments where devices are energy-constrained.

Non-orthogonal multiple access (NOMA) has emerged as a promising solution to enhance spectral efficiency and user connectivity in multi-user communication systems. Unlike conventional orthogonal multiple access (OMA) schemes, NOMA enables multiple users to share the same time–frequency resources by leveraging power-domain multiplexing and successive interference cancellation (SIC) techniques \cite{noma1, noma2}. This capability allows improved utilization of spectrum resources and supports the simultaneous transmission of multiple users with different channel conditions.
Meanwhile, wireless power transfer (WPT) and energy harvesting (EH) techniques enable devices to harvest energy from ambient or dedicated radio-frequency signals, reducing dependence on fixed power supplies and supporting long-term network sustainability \cite{boshkovska2015practical}. 

Despite these advancements, realizing the full potential of NOMA-based EH networks remains challenging, particularly in dynamic propagation environments and under practical hardware limitations. Most conventional architectures rely on fixed-position antennas at users or base station (BS) sides, which limits their ability to adapt to time-varying channel conditions and maintain reliable line-of-sight (LoS) connectivity especially in high-frequency bands such as millimeter-wave (mmWave) and terahertz (THz) spectra. These constraints can severely degrade system performance due to shadowing and non-line-of-sight (NLoS) propagation. Moreover, uplink NOMA resource optimization becomes highly complex due to the heterogeneity of users’ EH capabilities, battery states, and channel conditions \cite{10167828}. Therefore, new architectures that provide spatial adaptability and enhanced energy efficiency are essential for the next generation of wireless networks.

Recently, pinching-antenna (PA) arrays mounted on dielectric waveguides have emerged as an innovative and hardware-efficient solution to enhance spatial reconfigurability in wireless systems \cite{zeng2025energynoma, zeng2025sum, zeng2025robust, 11202497, 11300296}. Unlike conventional antenna arrays with fixed radiation elements, PA arrays allow the active radiation point to be dynamically reconfigured along the waveguide, enabling real-time adaptation to channel variations. This flexibility enhances beamforming capabilities and effective channel gains without increasing antenna count or hardware complexity. Consequently, PA-based transceivers can dynamically establish stronger LoS paths, mitigate blockage effects, and optimize spatial signal propagation in real time. When incorporated into wireless-powered NOMA networks, this spatial adaptability can simultaneously enhance harvested energy during the WPT phase and improve uplink channel conditions during information transmission. Hence, the integration of PA with WPT-assisted NOMA is not merely a combination of existing techniques, but a structurally motivated design aimed at overcoming the intrinsic limitations of fixed-antenna architectures.

Motivated by these advantages, this paper investigates an uplink NOMA network assisted by waveguide-mounted PA arrays under a harvest-then-transmit protocol. 
Distinct from existing PA-related studies that primarily focus on sum-rate maximization, beamforming design, or simplified resource allocation under ideal energy assumptions, this work explicitly considers the intrinsic coupling between spatial reconfigurability and energy-aware uplink transmission in EH-enabled NOMA systems.
In the proposed system, each user first harvests energy from the BS’s downlink transmission and then transmits data over a shared uplink channel. 
The primary objective is to maximize the system’s overall energy efficiency (EE) through joint optimization of antenna positioning, transmit power control, and time-switching ratios. The problem formulation accounts for realistic factors such as waveguide transmission losses, nonlinear EH characteristics, and uncertainties in user locations and battery states.
Due to the coupling between design variables and the stochastic nature of environmental and hardware parameters, the formulated optimization problem is non-convex and analytically intractable. To overcome this challenge, a deep reinforcement learning (DRL)-based optimization framework is proposed to autonomously learn efficient resource allocation strategies. The proposed DRL approach enables adaptive and data-driven decision-making without requiring explicit environmental models or perfect channel information.
Extensive simulation results demonstrate that the proposed PA-assisted uplink NOMA system achieves substantial improvements in both energy and spectral efficiency compared to conventional fixed-antenna and static resource allocation schemes.

\section{System Model}

\begin{figure}[t]
    \centering
    \subfloat[]{
    \includegraphics[width=0.75\linewidth,height=0.34\linewidth]{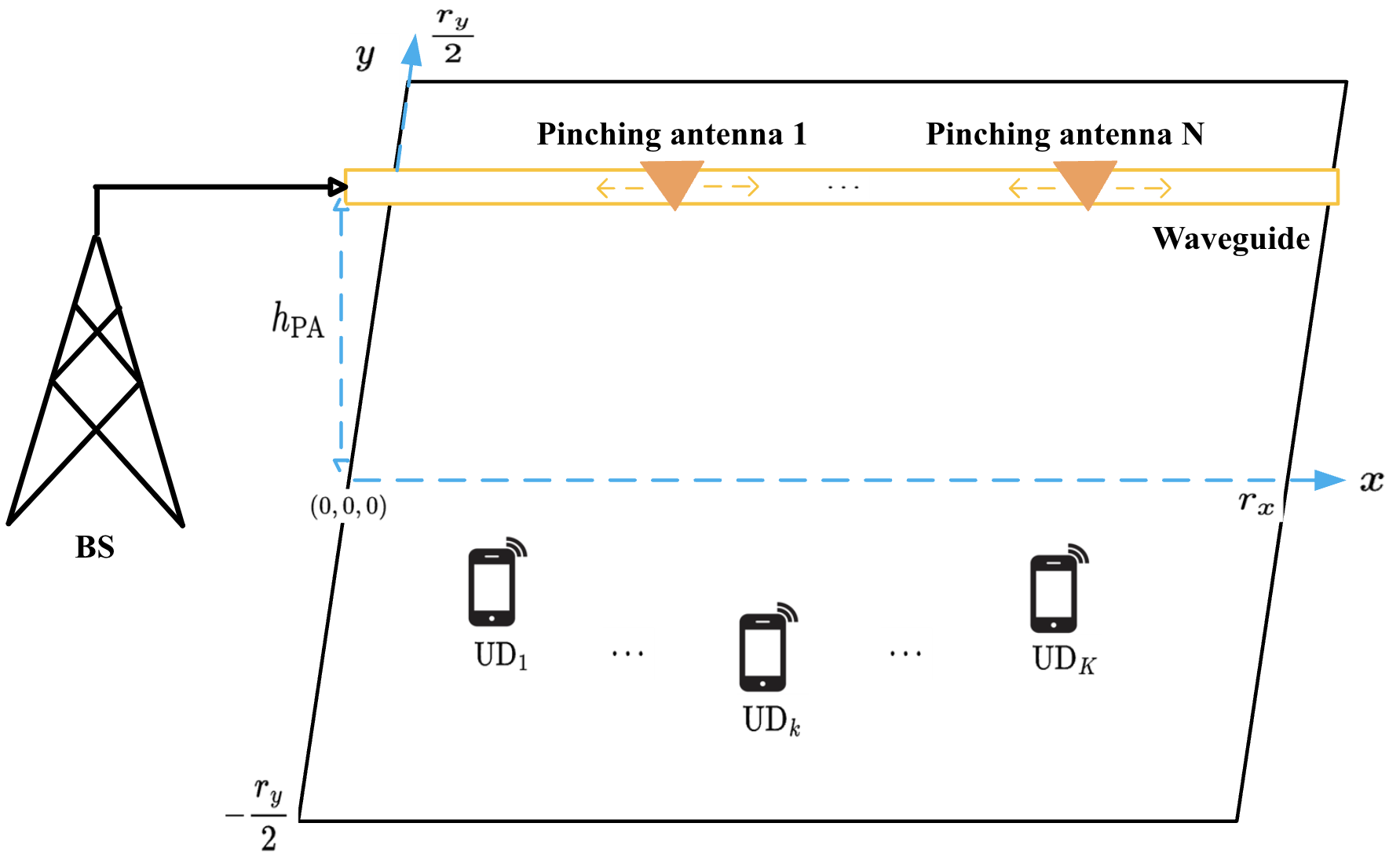}
        \label{fig:system_a}
    }\\[1.5mm] 
    \subfloat[]{
    \includegraphics[width=0.75\linewidth,height=0.12\linewidth]{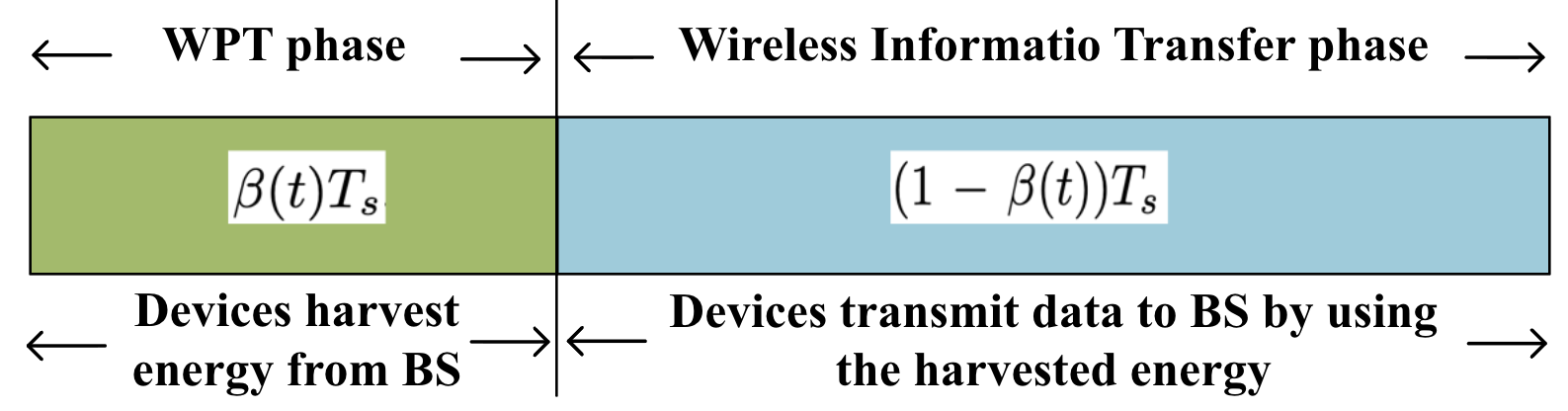}
        \label{fig:system_b}
    }
    \caption{System model and operation protocol of the proposed PA-assisted NOMA network under the harvest-then-transmit protocol.}
    \label{fig:system_model}
\end{figure}

We consider a multi-user uplink NOMA network, where single-antenna user devices (UDs), denoted by \( \text{UD}_k \), \( k \in \mathcal{K} \triangleq \{1,2,\ldots,K\} \), simultaneously transmit data to a BS. The BS is equipped with \(N\) PAs, denoted by \( \text{PA}_n \), \( n \in \mathcal{N} \triangleq \{1,2,\ldots,N\} \). Each PA is mounted on a dielectric waveguide of fixed height \(h_{{PA}}\) and length \(r_x\), and can be dynamically activated at a position $x_n$ along the waveguide, as illustrated in Fig.~\ref{fig:system_a}. This architecture provides flexible spatial control of the radiation points without increasing the antenna footprint.

The UDs are randomly distributed within a rectangular \(x\)–\(y\) service area of dimensions \(r_x \times r_y\) and remain stationary during each transmission interval. The position vectors of the \(n\)-th PA and the \(k\)-th UD are expressed as
$
\mathbf{d}_n = [x_n,\,0,\,h_{PA}]^{\mathrm{T}}$ and $
\mathbf{d}_k = [x_k,\,y_k,\,0]^{\mathrm{T}}
$ respectively, 
where \(x_k \in [0, r_x]\), \(y_k \in [-r_y/2, r_y/2]\), \(\forall k \in \mathcal{K}\), and \(\forall n \in \mathcal{N}\).  
The overall channel gain between the activated PAs and \( \text{UD}_k \) is characterized by the coherent superposition of the individual channel contributions from all PAs \cite{li2025sum}. Let \( \mathbf{x} = [x_1, x_2, \ldots, x_N] \) denote the PA activation vector. Accordingly, the resulting composite channel gain is given by
\begin{equation}
H_k(\mathbf{x}) = \sum_{n=1}^{N} h_{k}(x_n),
\end{equation}
where \(h_k(x_n)\) represents the end-to-end complex channel coefficient between $\mathrm{UD}_k$ and the $n$-th PA at position $x_n$. This coefficient incorporates the effects of waveguide propagation, PA–waveguide coupling, and free-space transmission, thereby capturing both electromagnetic and spatial characteristics of the system. We adopt the proportional power consumption model \cite{wang2025modeling} and the waveguide transmission loss model \cite{wang2025antenna} to capture practical deployment considerations. These considerations ensure that the proposed system captures both the hardware-level and propagation-level impacts on the achievable network performance.

Assuming isotropic radiation for each PA, the channel coefficient can be expressed as
\begin{equation}
\begin{aligned}
&h_{k}(x_n)=\\ 
& \underbrace{\frac{\lambda}{4\pi  \|\mathbf{d}_k - \mathbf{d}_n\|}  
e^{-j \tfrac{2\pi}{\lambda} \|\mathbf{d}_k - \mathbf{d}_n\|}}_{\text{free-space propagation}}
\times \underbrace{\psi_n^{1/2} \varsigma_n^{1/2} e^{-j \tfrac{2\pi}{\lambda_g} |\mathbf{d}_0 - \mathbf{d}_n|}}_{\text{\makecell{waveguide propagation\\ and coupling}}},
\end{aligned}
\end{equation}
where \(\lambda\) and \(\lambda_g = \lambda / n_{\mathrm{eff}}\) denote the free-space and guided wavelengths, respectively. The reference point \(x_0\) denotes the excitation location of the waveguide feed, and \(n_{\mathrm{eff}}\) is the effective refractive index of the guided medium.

The power allocation and waveguide loss coefficients for the \(n\)-th PA are respectively given by
$\psi_n = \delta^2 (1 - \delta^2)^{n-1}$ and $ 
\varsigma_n = 10^{-\tfrac{\mu |\mathbf{d}_0 - \mathbf{d}_n|}{10}}$, where \(\delta = \sin(\kappa L)\), $\kappa$ denotes the PA–waveguide coupling coefficient, $L$ is the coupling length, and $\mu$ represents the dielectric attenuation factor \cite{li2025sum, 10981775}.

We consider communication over \(T\) time slots, each of duration \(T_s\). To avoid interference between power transfer and data transmission, a two-phase harvest-then-transmit protocol is employed, as illustrated in Fig.~\ref{fig:system_b}. During each time slot $t$, the BS performs downlink WPT for a fraction $\beta^{(t)} T_s$ of the total duration, where $\beta^{(t)} \in [0,1]$ denotes the time-switching ratio. The harvested energy is stored in a rechargeable battery and subsequently used during the remaining \((1-\beta^{(t)}) T_s\) for uplink wireless information transmission, in which the UDs transmit data to the BS.

In Phase~I, the harvested energy at \(\mathrm{UD}_k\) during time slot $t$ follows a practical nonlinear EH model~\cite{li2025sum,boshkovska2015practical}, which captures the nonlinear response of the rectifier circuit as 
\begin{equation}
	E_k^{(t)} = \beta^{(t)} T_s \frac{\Upsilon - \iota \Omega}{1 - \Omega},
\end{equation}
where \(\Upsilon = \frac{\iota}{1 + e^{-a (p_{0} |H_k(\mathbf{x})|^2 - b)}}\), \(\Omega = \frac{1}{1 + e^{ab}}\), and $a$, $b$, and $\iota$ are hardware-dependent parameters characterizing the sensitivity, activation threshold, and saturation level of the rectifying circuit, respectively. Specifically, $a$ controls the steepness of the nonlinear charging curve, $b$ determines the minimum input RF power required to effectively activate the EH process, and $\iota$ represents the maximum harvested power achieved at saturation. Here, $p_0$ denotes the BS transmit power used for WPT.

In Phase~II, all UDs simultaneously transmit their data using power-domain NOMA. The received superimposed signal is decoded at the BS using SIC. To guarantee fairness among users, the UDs are ordered
in descending order according to their effective channel gains,
i.e., $|H_1(\mathbf{x})|^2 \ge  |H_2(\mathbf{x})|^2 \ge \cdots \ge |H_K(\mathbf{x})|^2$, and are decoded subsequently following this order using SIC. This channel-gain-based decoding order is adopted to ensure reliable SIC and analytical tractability. Since transmit power depends on harvested energy, which is channel-dependent, the received signal strength inherently reflects both channel quality and energy availability. Dynamic permutation-based ordering is left for future work due to its combinatorial complexity. The achievable uplink rate of $\mathrm{UD}_k$ in slot $t$ is given by~\cite{sheikhzadeh2021ai}
\begin{equation}
R_k^{(t)} = (1-\beta^{(t)}) \log_2 \left( 1 + \frac{p_k^{(t)} |H_k(\mathbf{x})|^2}{\sum_{j=k+1}^{K} p_j^{(t)} |H_j(\mathbf{x})|^2 + \sigma^2} \right),
\end{equation}
where \(p_k^{(t)}\) is the transmit power of $\mathrm{UD}_k$, and \(\sigma^2\) represents the additive white Gaussian noise (AWGN) power at the BS.

The remaining battery energy is updated at the end of each time slot as~\cite{sheikhzadeh2021ai}
\begin{equation}
	B_k^{(t+1)} = \min \{ B^{\max}_{k},\max\{ B_k^{(t)}  + \eta_k E_k^{(t)}  - E^{{(t)}}_{k,\mathrm{cons}}, 0\}    \},
\end{equation}
where \(\eta_k \in (0,1]\) denotes the storage efficiency, $B_k^{\max}$ is the maximum battery capacity, and $E_{k,\mathrm{cons}}^{(t)}$ is the total consumed energy in slot $t$, defined as
\begin{equation}
E_{k,\mathrm{cons}}^{(t)} = (1 - \beta^{(t)}) T_s p_k^{(t)} + E_k^{\mathrm{f}},
\label{eq:consumed}
\end{equation}
where $E_k^{\mathrm{f}}$ represents the fixed circuit energy consumption. The energy feasibility constraint requires
\begin{equation}
E_{k,\mathrm{cons}}^{(t)} \leq B_k^{(t)} + \eta_k E_k^{(t)},
\label{eq:constraint}
\end{equation}
ensuring that the total consumed energy never exceeds the available stored and harvested energy.

To capture practical deployment scenarios, we consider two types of uncertainties: imperfect knowledge of user battery levels and imperfect knowledge of user spatial locations. These uncertainties are treated separately as follows.

\subsubsection{Battery Level Uncertainty}
The actual battery level of $\mathrm{UD}_k$ at time slot $t$ is expressed as
\begin{equation}
B_k^{(t)} = \hat{B}_k^{(t)} + \Delta_{b,k}^{(t)}, \quad |\Delta_{b,k}^{(t)}| \le \Delta_{b,0},
\end{equation}
where $\hat{B}_k^{(t)}$ denotes the estimated battery level, $\Delta_{b,k}^{(t)}$ is the estimation error, and $\Delta_{b,0}$ is the maximum allowable deviation in battery estimation. The battery state error is assumed to be uniformly distributed within this bounded interval, reflecting uncertainties due to sensing inaccuracies, quantization effects, or delayed energy reporting.

\subsubsection{Spatial Location Uncertainty}
Similarly, the actual location of $\mathrm{UD}_k$ at time slot $t$ is modeled as
\begin{equation}
\mathbf{r}_k^{(t)} = \hat{\mathbf{r}}_k^{(t)} + \boldsymbol{\Delta}_{r,k}^{(t)}, \quad \|\boldsymbol{\Delta}_{r,k}^{(t)}\| \le \Delta_{r,0},
\end{equation}
where $\hat{\mathbf{r}}_k^{(t)}$ denotes the estimated location, $\boldsymbol{\Delta}_{r,k}^{(t)}$ is the estimation error, and $\Delta_{r,0}$ defines the maximum admissible deviation. The spatial error is modeled as a bounded Gaussian random vector with zero mean and variance $\sigma_r^2$.

Battery and location uncertainties follow different distributions to reflect their underlying physical characteristics. Battery measurement errors are inherently bounded by the sensor resolution and are equally likely within that range, hence a uniform model is appropriate. In contrast, location estimation errors arise from multiple independent stochastic factors, naturally resulting in a Gaussian distribution. 

This model captures both spatial uncertainty and imperfect energy state information, which are critical in adaptive resource allocation in dynamic PA-assisted NOMA systems.

For clarity and ease of reference, the key symbols and parameters used throughout the paper are summarized in Table~\ref{tab:symbols}.

\begin{table}[t]
\caption{Summary of Key Symbols and Parameters}
\centering
\begin{tabular}{ll}
\toprule
Symbol & Description \\
\midrule
$K$ & Number of user devices (UDs) \\
$N$ & Number of pinching antennas (PAs) \\
$x_n$ & Activation position of the $n$-th PA \\
$\mathbf{x}$ & PA position vector \\
$H_k(\mathbf{x})$ & Composite channel gain of $\mathrm{UD}_k$ \\
$p_k^{(t)}$ & Transmit power of $\mathrm{UD}_k$ at slot $t$ \\
$p_0$ & BS transmit power for WPT \\
$\beta^{(t)}$ & Time-switching ratio for EH \\
$E_k^{(t)}$ & Harvested energy of $\mathrm{UD}_k$ \\
$B_k^{(t)}$ & Battery level of $\mathrm{UD}_k$ \\
$\eta_k$ & Energy storage efficiency \\
$R_k^{(t)}$ & Achievable uplink rate of $\mathrm{UD}_k$ \\
$R_{\min}$ & Minimum rate requirement \\
$\sigma^2$ & Noise power at the BS \\
$\mu$ & Waveguide attenuation coefficient \\
$\psi_n$ & Power allocation coefficient of $n$-th PA \\
$\varsigma_n$ & Waveguide transmission loss factor \\
\bottomrule
\end{tabular}
\label{tab:symbols}
\end{table}

\subsection{Problem Formulation}

In conventional wireless systems, sum-rate maximization often requires allocating the maximum transmit power to users, which may lead to energy-inefficient operation, particularly in resource-constrained networks. To balance spectral efficiency with sustainable energy usage, this work focuses on maximizing the EE, defined as the ratio of the aggregate achievable rate to the total power consumption, which
includes both the fixed circuit power and the dynamic transmit power \cite{333333, zeng2025energynoma}.

Accordingly, the EE maximization problem is formulated as a joint optimization of the transmit power vector $\mathbf{p} = [p_1, \dots, p_K]$, the PA position vector $\mathbf{x} = [x_1, \dots, x_N]$, and the time-switching ratio $\beta^{(t)}$ that determines the fraction of time allocated for EH versus information transmission. The problem is expressed mathematically as:
\begin{align}
\small
\max_{\beta^{(t)},\,\mathbf{p},\,\mathbf{x}} \quad 
& \sum_{t=1}^{T}\left( \frac{ \sum_{k=1}^{K} R_k^{(t)}}{P_f+\sum_{k=1}^{K} p_k^{(t)}} \right) \label{eq:EEmax_corrected} \\
\text{s.t.} \quad  
& C_1: \mathbf{x}_n \in [0, r_x], \quad \forall n \in \{1,\dots,N\}, \label{const:Xrange} \\
& C_2: |\mathbf{x}_n - \mathbf{x}_{n-1}| \geq d_{\min}, \ \forall n\in \left\{ 2, ..., N \right\}, \label{const:Xspacing}  \\
& C_3: \beta^{(t)} \in [0,1], \quad \forall t, \label{const:beta} \\
& C_4: R_k^{(t)} \geq R_{\min}, \quad \forall k,t, \label{const:Rmin}\\
& C_5: E_{k,{\mathrm{cons}}}^{(t)} \leq B_k^{(t)} + \eta_k E_k^{(t)},
\\
& C_6: (8) \ \text{and} \ (9). 
\end{align}
Here, constraints $C_1$ and $C_2$ ensure that each PA is located within the valid operational range of the waveguide while maintaining a minimum spacing $d_{\min}$ between adjacent antennas to avoid interference or mutual coupling. $C_3$ enforces a feasible time-switching ratio, guaranteeing a valid division between energy harvesting and information transmission periods. $C_4$ ensures that each user meets the minimum data rate requirement, while $C_5$ guarantees that the energy consumed by user $k$ during time slot $t$ does not exceed the sum of its available battery energy and the harvested energy. Finally, $C_6$ incorporates the effects of uncertainties in both battery levels and spatial locations, ensuring robust system performance under imperfect knowledge.

The joint optimization problem is inherently non-convex due to the coupling among key system parameters. This complexity is further intensified by battery-level and spatial-location uncertainties. Battery uncertainty makes the energy causality constraint time-varying, while spatial uncertainty perturbs the effective channel gains in both harvested energy and rate expressions. Consequently, the objective function and constraints become stochastic and dynamically coupled, increasing solution intractability. To address these challenges and enable adaptive decision-making in uncertain and dynamic environments, this work adopts a DRL–based solution framework. The proposed approach leverages data-driven learning to jointly optimize network parameters in real time, aiming to achieve a near-optimal balance between EE and communication performance.

	\section{Proposed DRL Algorithm}
	
	To efficiently solve the non-convex and dynamically coupled optimization problem, we deploy a DRL agent at the BS. The agent learns an adaptive policy to jointly optimize user transmit power, PA activation positions, and the time allocation between EH and uplink information transmission. 
The optimization is formulated within a Markov decision process (MDP) framework, where the agent autonomously controls the PA activation positions, user transmit powers, and time-switching ratios to maximize EE while satisfying system constraints. The definitions of state, action, and reward are as follows:    
	\begin{itemize}
		\item \textbf{State Space}: The system state at each time step includes the remaining battery energy of all UDs and their spatial locations within the service area, represented as $\boldsymbol{{s}}_t = \{ [B_1^{(t)}, \dots, B_K^{(t)}], [d_1, \dots, d_K] \}$.

\item \textbf{Action Space}: The action at each time step consists of the transmit power of all UDs, the positions of all PAs, and the time-switching ratio for EH, represented as $\boldsymbol{a}_t = \left\{ [p_1^{(t)}, \dots, p_K^{(t)}], [{x}_1, {x}_2, \dots, {x}_N], [\beta^{(t)}] \right\}$. All physical constraints are enforced through action space design and environment-level feasibility checks. The actor outputs are mapped onto feasible ranges, and transmit power is bounded by the available battery energy during state transitions. Thus, the DRL agent operates within a feasible action space.

\item \textbf{Reward function}: To maximize EE while satisfying minimum rate requirements, the reward function is defined as
		\begin{equation}
			r(\boldsymbol{s}_t, \boldsymbol{a}_t) = 
			\begin{cases}
				r_p, & \text{if } R_k^{(t)} \leq R_{\min}, \\
			 \frac{ \sum_{k=1}^{K} R_k^{(t)}}{P_f+\sum_{k=1}^{K} p_k^{(t)}}, & \text{otherwise},
			\end{cases}
		\end{equation}
where $r_p$ is a penalty term that needs to be tuned during the simulation to achieve optimal convergence behavior. It is set as a sufficiently large negative constant to penalize violations of the minimum rate constraint, with its magnitude chosen relative to the achievable EE under feasible conditions.

	\end{itemize}

	Given the continuous nature of the action space in our problem, conventional model-free value-based DRL algorithms such as deep Q-networks (DQN) are unsuitable. We therefore adopt a policy-gradient approach, specifically the deep deterministic policy gradient (DDPG) algorithm, which is well suited for continuous action spaces \cite{sheikhzadeh2021ai, 9448276}. DDPG is particularly suitable for our problem because it efficiently learns deterministic policies in high-dimensional continuous action spaces, providing stable and sample-efficient training for joint optimization of coupled resource allocation parameters. Although DDPG is adopted as the learning backbone, its implementation is tailored to the considered system. The state design captures battery and spatial uncertainties, the actions are normalized and projected to satisfy physical constraints, and the reward is aligned with the fractional EE objective.
    Our DDPG framework consists of four neural networks. The actor network, denoted by \( \pi_{\phi} \) with parameters \( \phi \), generates actions \( \boldsymbol{a}_t \) from the current state \( \boldsymbol{s}_t \), incorporating exploration noise as \( \boldsymbol{a}_t = \pi_{\phi}(\boldsymbol{s}_t) + \xi \). The critic network, parameterized by \( \theta \), evaluates state-action pairs and computes the corresponding Q-values \( Q_{\theta}(\boldsymbol{s}_t, \boldsymbol{a}_t) \), which estimates the expected cumulative reward. Both actor and critic are implemented as fully connected neural networks with multiple hidden layers and ReLU activations. The hidden-layer widths are selected to provide sufficient representation capacity while maintaining stable training. For stable neural network training, all state and action variables are linearly normalized according to their feasible ranges before being fed into the actor and critic networks. The actor outputs normalized actions, which are subsequently mapped back to their corresponding physical domains prior to system implementation. This scaling mitigates gradient instability arising from heterogeneous variable magnitudes and improves convergence behavior. In the critic, the concatenated state–action input is followed by layer normalization, and soft target updates are used to improve training stability. The framework aims to learn an optimal policy to maximize cumulative 
    expected rewards, defined as
    \begin{equation}
		\pi^* = \arg \max_{\pi} \mathbb{E}_{\boldsymbol{s}_t, \boldsymbol{a}_t} \left[ \sum_{t=0}^{\infty} r(\boldsymbol{s}_t, \boldsymbol{a}_t) \right].
	\end{equation}
	The actor network is updated by the gradient of the objective function \( J(\phi) \), given by
	\begin{equation}
		\nabla_{\phi} J(\phi) = \mathbb{E} \left[ \nabla_{\boldsymbol{a}_t} Q_{\theta_1}(\boldsymbol{s}_t, \boldsymbol{a}_t) \bigg|_{\boldsymbol{a}_t=\pi_{\phi}(\boldsymbol{s}_t)} \nabla_{\phi} \pi_{\phi}(\boldsymbol{s}_t) \right].
	\end{equation}
	Concurrently, the critic network is updated to minimize the error between its predictions and the target values, defined as
	\begin{equation}
		Y_t = r_t + \gamma Q_{\theta_i'}(\boldsymbol{s}_{t+1}, \pi_{\phi'}(\boldsymbol{s}_{t+1}) + {\xi}).
		\label{eq:target}
	\end{equation}

 The proposed DDPG method is described in Algorithm \ref{algorithm}.

	\RestyleAlgo{ruled}
	\SetNlSty{textbf}{}{:} 
	\begin{algorithm}
		\SetAlgoLined
		\textbf{Initialize}: Experience replay buffer \( M \), mini-batch size \( H \), actor network \( \pi_{\phi} \), critic network \( Q_{\theta} \) with random values, and create the target networks by setting \( \theta' \leftarrow \theta \) and \( \phi' \leftarrow \phi \).
		
		\textbf{Set}: Maximum episodes \(E\) and episode length \(T\).\\
		
		\For{each episode $e:E{}$}{%
			Initialize state \(\boldsymbol{s}_0\) and exploration noise process \(\xi\);\\
			\For{$t = 1:T$}{%
				Observe current state \(\boldsymbol{s}_t\) from the environment;\\
				Generate the action $\boldsymbol{a}_t = \pi_{\phi}(\boldsymbol{s}_t) + \xi$ and apply the required reshaping; \\ 
				Compute $r_t$ using (17);\\
				Observe the next state $\boldsymbol{s}_{t+1}$;\\
				Store the transition $(\boldsymbol{s}_t, \boldsymbol{a}_t, r_t, \boldsymbol{s}_{t+1})$ into the buffer $M$;\\ 
			}
			Sample a mini-batch of size \(H\) from \(M\);\\
			Evaluate the target values $Y_t$ using \eqref{eq:target};\\
			Update actor and critic parameters using the Adam optimizer;\\
			Perform a soft update of the target networks using the update coefficient \(\tau \in [0,1]\):
			\[
			\phi' \leftarrow \tau \phi + (1 - \tau)\phi', \quad
			\theta' \leftarrow \tau \theta + (1 - \tau)\theta'
			\]}	
		\caption{The DDPG Algorithm}
		\label{algorithm}
	\end{algorithm}

\section{Simulation Results}

This section presents numerical results to validate the performance of the proposed PA–assisted uplink NOMA network under the harvest-then-transmit protocol. The algorithm is initialized with the following parameters \cite{zeng2025energynoma, zeng2025sum, li2025sum, pakravan2025fluid}. The
carrier frequency is set to $28$~GHz, the thermal noise power is $\sigma^2 = -90$~dBm, the waveguide attenuation coefficient is $\mu = 0.5~\mathrm{dB/m}$, and the waveguide length equals the antenna spacing, $L = {r_x}$. User locations are uniformly distributed within a rectangular region of size $r_x = 60$~m and $r_y = 20$~m. The BS is equipped with $N = 3$ PAs serving $K = 3$ UDs, with initial antenna offset $X_0 = 0.5\lambda$ and total waveguide length $X = 8\lambda$. 
At the beginning of each training episode, the initial battery level of each user is randomly drawn from a uniform distribution over the feasible range $[0, B_k^{\max}]$. The initial PA positions are uniformly distributed along the waveguide while satisfying the minimum spacing constraint and feasible location bounds. The Adam optimizer is adopted with a learning rate of $5\times10^{-4}$, batch size of 64, replay buffer size of $10^4$, soft update rate of $0.001$, and discount factor $\gamma = 0.9$. The average reward over 100 episodes at episode $e$ is computed as \({R_{\text{avg}}(e)} = \frac{1}{100} \sum_{i=e-100}^{e} R_i\), where \( R_i \) denotes the mean reward obtained in episode \( i \).

To assess the effectiveness of the proposed scheme, we compare its performance with the following benchmarks:

\begin{itemize}
    \item \textbf{Constrained Continuous Optimization:} In this configuration, each PA is capable of continuously moving within a predefined range along the dielectric waveguide, following a sequential pattern with equal displacement intervals for all $N$ elements.

\item \textbf{Discrete PA Positioning:} Each PA is restricted to a predefined set of discrete locations along the waveguide, and optimization is performed over this fixed grid, yielding a semi-practical configuration with limited adaptability.

    \item \textbf{Fixed PA Positions:} 
    All PAs are uniformly placed along the waveguide and remain static during operation. This configuration corresponds to a conventional fixed-antenna system and serves as the baseline.

\item \textbf{PA-assisted OMA:}
{In this benchmark, the same system model is considered but users employ OMA, where the uplink transmission time is equally divided among users and only one user transmits at a time.} 
    
\end{itemize}

\begin{figure*}[]
\centering
\subfloat[]{\includegraphics[width=0.33\linewidth,height=0.2\linewidth]{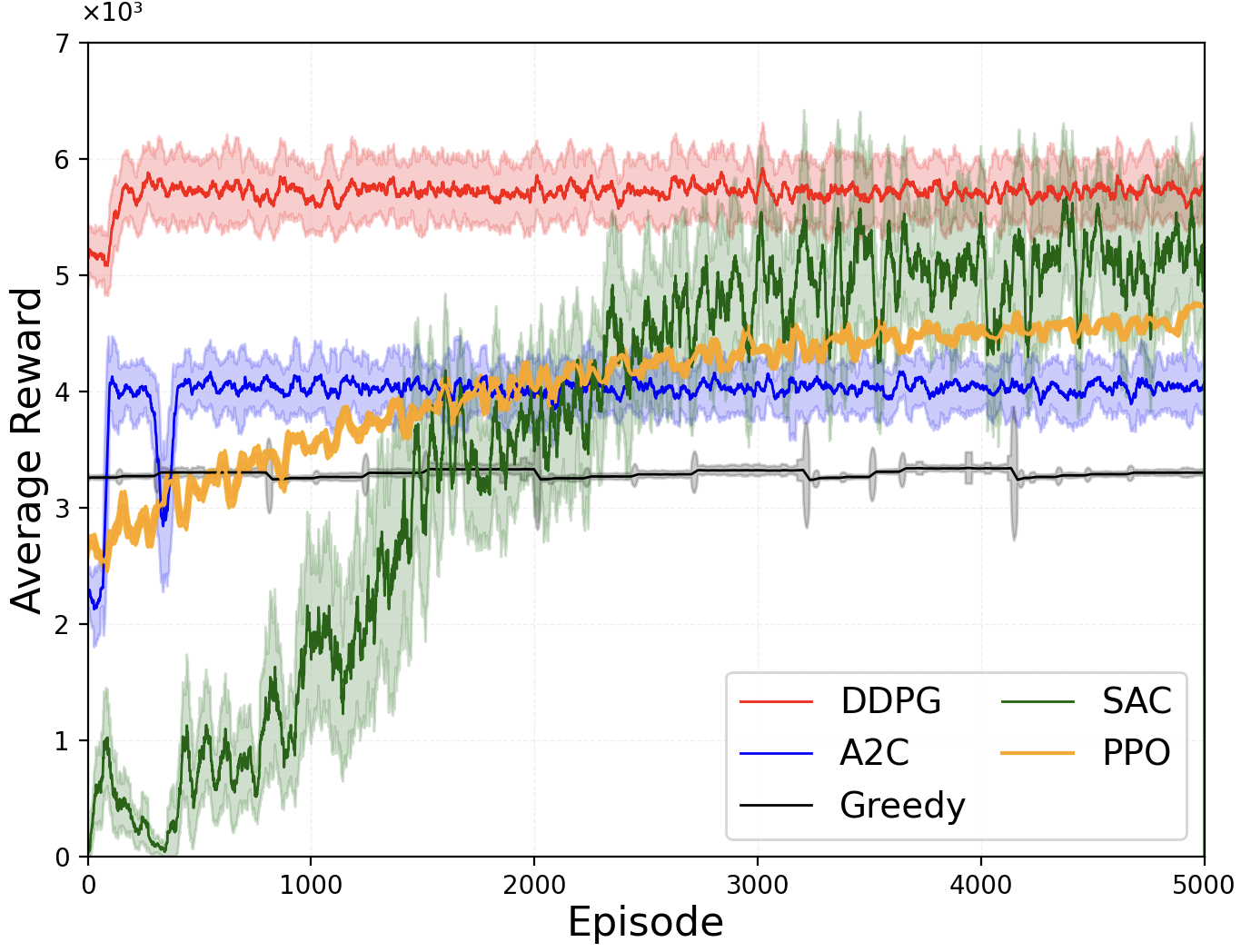}\label{fig:convergence}}
\hfil
\subfloat[]{\includegraphics[width=0.33\linewidth,height=0.2\linewidth]{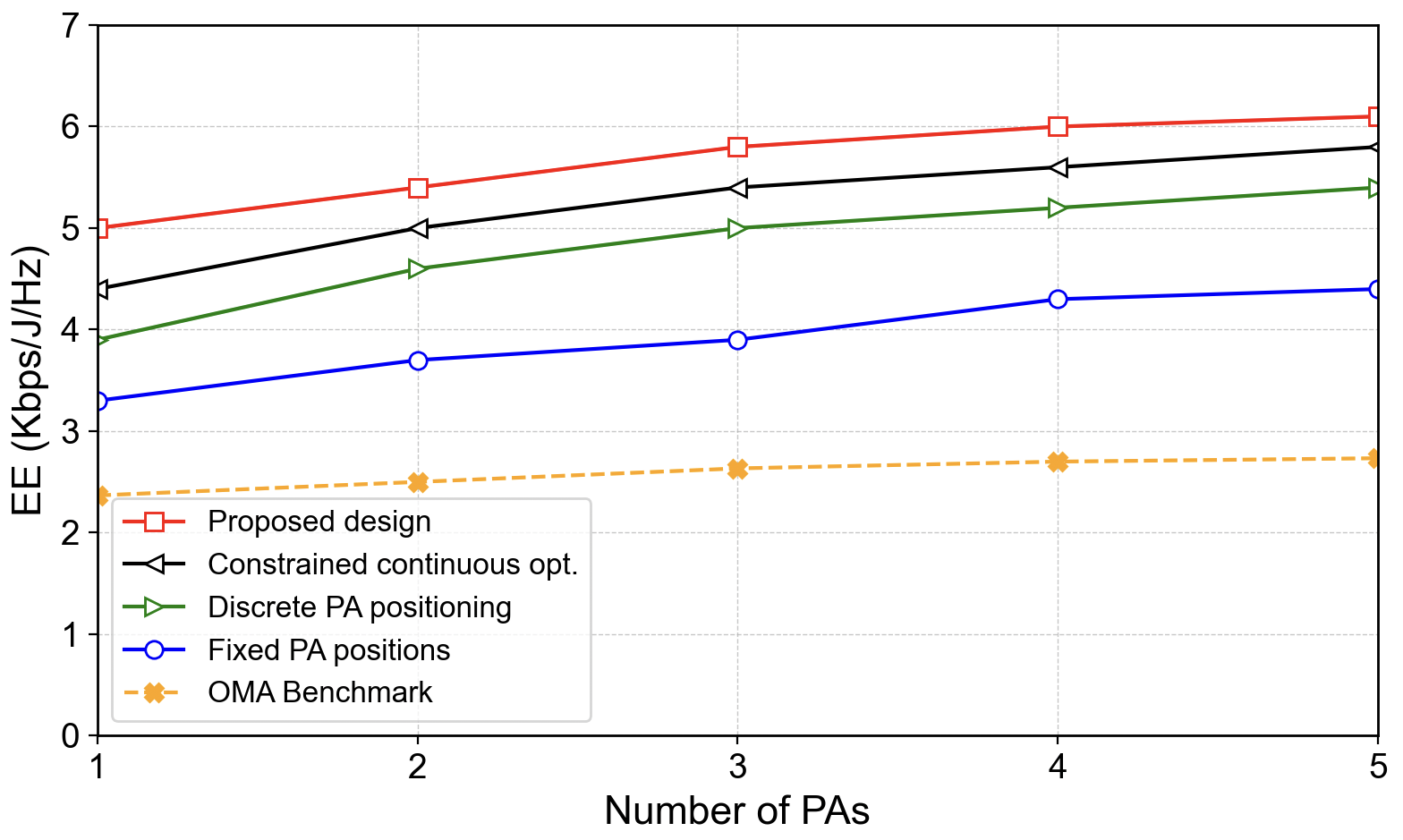}\label{fig:EE_PA}}
\hfil
\subfloat[]{\includegraphics[width=0.33\linewidth,height=0.2\linewidth]{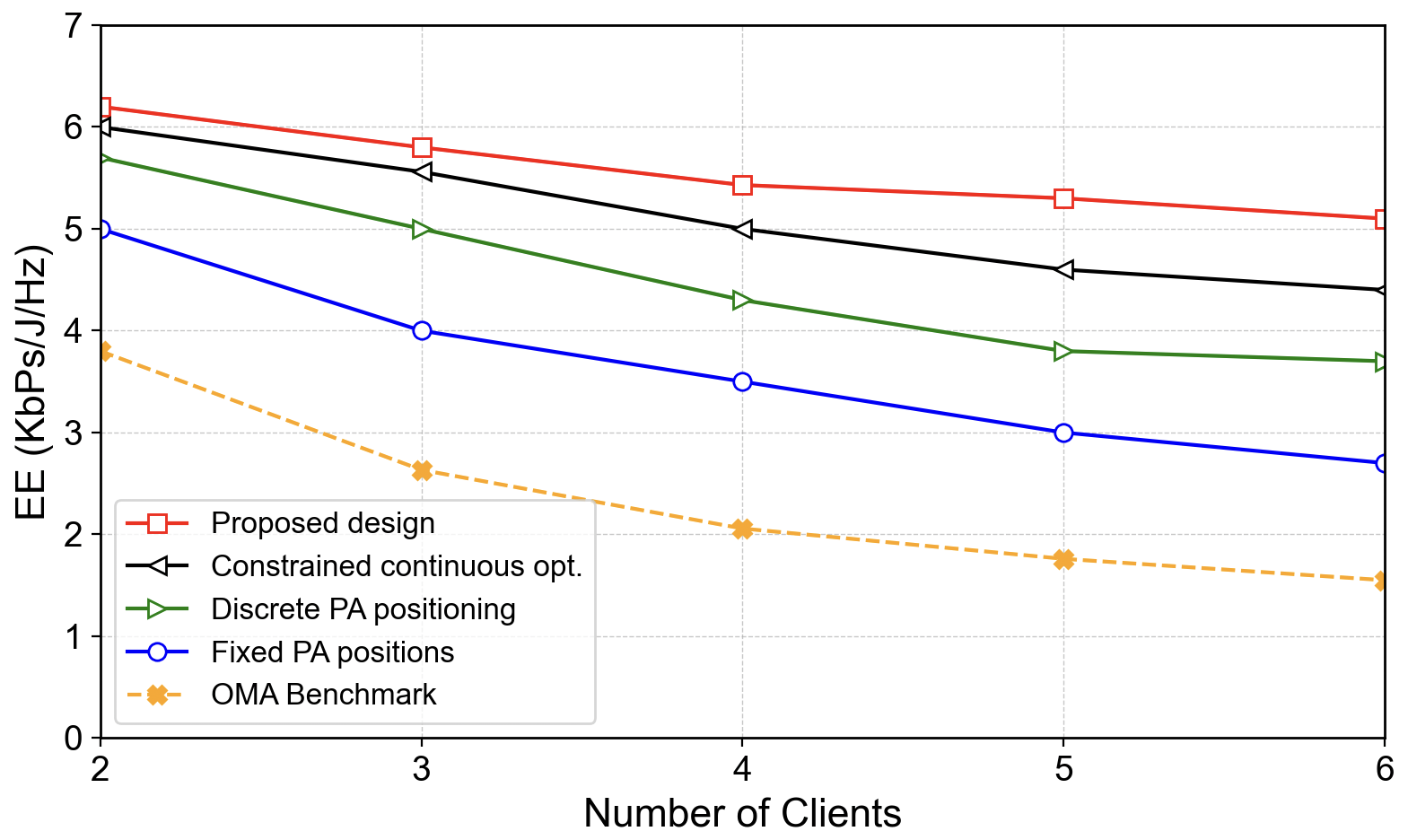}\label{fig:EE_clients}}
\caption{Performance evaluation of the proposed DRL-enabled adaptive PA positioning framework: 
(a) convergence of average reward; 
(b) EE versus the number of PAs; and 
(c) EE versus the number of clients.}
\label{fig:simulation_results}
\end{figure*}

Fig.~\ref{fig:convergence} presents the convergence behavior of different DRL algorithms \cite{pakravan2025fluid, liu2023evaluation, 11111711} in terms of the average reward over training episodes. The early fluctuations in the reward curves are primarily caused by exploration noise and the agent’s untrained policy, which gradually stabilizes as the DRL agent learns effective actions. The solid curves correspond to the average accumulated rewards, while the shaded regions depict the standard deviation across multiple runs, thereby reflecting the stability and consistency of each method. The results indicate that the DDPG algorithm consistently achieves the highest average rewards throughout the training process. Moreover, it exhibits significantly lower variability compared to the baseline methods.The superior performance of DDPG is mainly attributed to its deterministic policy structure, which is well suited for the tightly coupled and constraint-sensitive EE maximization problem. In contrast, SAC introduces entropy regularization for exploration, while A2C and PPO rely on stochastic policy updates, which may lead to higher variance and slower convergence in constrained continuous resource allocation problems.

Fig.~\ref{fig:EE_PA} illustrates the system EE versus the number of PAs. The proposed adaptive design yields a steady increase in EE as more PAs are deployed, demonstrating its capability to efficiently exploit the additional spatial degrees of freedom introduced by multiple pinching elements. The initial increase in EE arises from enhanced beamforming flexibility and spatial diversity, which enable more effective signal alignment and interference mitigation. Nevertheless, as the number of PAs continues to grow, the marginal gain in EE gradually diminishes due to the rising circuit power consumption and the coupling effects among closely spaced pinching elements.

Fig.~\ref{fig:EE_clients} depicts the EE performance as the number of users
increases. The results show a monotonic decline in EE due to intensified inter-user interference, elevated resource contention, and cumulative circuit power consumption. The initial decrease is pronounced, indicating that the system is highly sensitive to the addition of the first few clients, whereas the rate of decline gradually diminishes in denser network scenarios, reflecting the saturation of available spatial and spectral resources. Notably, the proposed adaptive design consistently achieves higher EE compared to baseline approaches.

\section{Conclusion}

This paper has investigated an uplink NOMA system enhanced by PA arrays operating under a harvest-then-transmit protocol. By dynamically adjusting the radiation points along dielectric waveguides, the proposed PA-assisted design enables adaptive beamforming and improved channel gains without additional hardware complexity. A joint DRL-based optimization framework for antenna positioning, transmit power, and time-switching ratio was formulated to maximize system EE, and a DRL-based approach was developed to efficiently handle the resulting non-convex problem under uncertainties in user locations and battery states. Numerical results demonstrated that the proposed scheme significantly outperforms conventional fixed-antenna NOMA networks in terms of EE.

	\bibliographystyle{IEEEtran}
	\bibliography{refpinching}

@article{noma1,
  author = {Ming Zeng and others},
  title     = {Energy-Efficient Power Allocation in Uplink mmWave Massive {MIMO} With {NOMA}},
  journal   = {IEEE Trans. Veh. Technol.},
  volume    = {68},
  number    = {3},
  pages     = {3000-3004},
  month     = {Mar.},
  year      = {2019}
}

@article{noma2,
  author    = {S. M. R. Islam and others},
  title     = {Resource Allocation for Downlink {NOMA} Systems: Key Techniques and Open Issues},
  journal   = {IEEE Wireless Commun.},
  volume    = {25},
  number    = {2},
  pages     = {40--47},
  month     = {Apr.},
  year      = {2018}
}

@article{boshkovska2015practical,
  author    = {E. Boshkovska and others},
  title     = {Practical Non-Linear Energy Harvesting Model and Resource Allocation for {SWIPT} Systems},
  journal   = {IEEE Commun. Lett.},
  volume    = {19},
  number    = {12},
  pages     = {2082--2085},
  month     = {Dec.},
  year      = {2015}
}

@article{10167828,
  author    = {K. Agrawal and others},
  title     = {Performance of a Multiuser Cooperative {IoT} {NOMA} Network With Battery-Assisted Energy Harvesting},
  journal   = {IEEE Trans. Ind. Informat.},
  volume    = {20},
  number    = {2},
  pages     = {2307--2319},
  month     = {Feb.},
  year      = {2024}
}

@article{zeng2025energynoma,
  author    = {M. Zeng and others},
  title     = {Energy-Efficient Resource Allocation for {NOMA}-Assisted Uplink Pinching-Antenna Systems},
journal   = {IEEE Wireless Commun. Lett.},
  volume    = {14},
  number    = {11},
  pages     = {3695 - 3699},
  month     = {Nov.},
  year      = {2025}
}

@article{zeng2025sum,
  author    = {Y. Ai and others},
  title     = {Delay Minimization in Pinching-Antenna-enabled {NOMA-MEC} Networks},
  journal   ={IEEE Commun. Lett.},
  volume    = {30},
  number    = {},
  pages     = {962 - 966},
  month     = {Jan.},
  year      = {2026}
}

@article{zeng2025robust,
  author    = {K. Cao and others},
  title     = {Performance Analysis of Wireless-Powered Pinching Antenna Systems},
  journal   = {arXiv preprint arXiv:2511.03401},
  month     = {Nov.},
  year      = {2025}
}

@article{li2025sum,
  author    = {Y. Li and others},
  title     = {Pinching Antenna-Aided Wireless Powered Communication Networks},
  journal   = {IEEE Wireless Commun. Lett.},
 volume    = {15},
  number    = {},
  pages     = {255 - 259},
  month     = {Oct.},
  year      = {2025}
}

@article{pakravan2025fluid,
  author    = {S. Pakravan and others},
  title     = {Fluid Antenna-Assisted Uplink {NOMA} Networks under Imperfect {SIC}},
  journal   = {IEEE Trans. Veh. Technol.},
  month     = {Jan.},
 year={2026},
  volume={75},
  number={1},
  pages={1689-1694}}

@article{wang2025modeling,
  author    = {Z. Wang and others},
  title     = {Modeling and Beamforming Optimization for Pinching-Antenna Systems},
journal={IEEE Trans. Commun.}, 
  year={2025},
   month     = {Dec.},
  volume={73},
  number={12},
  pages={13904-13919}}

@article{wang2025antenna,
  author    = {K. Wang and others},
  title     = {Antenna Activation for {NOMA}-Assisted Pinching-Antenna Systems},
  journal   = {IEEE Wireless Commun. Lett.},
  month     = {May.},
year={2025},
  volume={14},
  number={5},
  pages={1526-1530}
}

@article{sheikhzadeh2021ai,
  author    = {S. Sheikhzadeh and others},
  title     = {{AI}-Based Secure {NOMA} and Cognitive Radio-Enabled Green Communications: Channel State Information and Battery Value Uncertainties},
  journal   = {IEEE Trans. Green Commun. Netw.},
  volume    = {6},
  number    = {2},
  pages     = {1037--1054},
  month     = {Dec.},
  year      = {2021}
}

@article{333333,
  author    = {M. Zeng and others},
  title     = {Spectral and Energy-Efficient Resource Allocation for Multi-Carrier Uplink {NOMA} Systems},
  journal   = {IEEE Trans. Veh. Technol.},
  volume    = {68},
  number    = {9},
  pages     = {9293--9296},
  month     = {Mar.},
  year      = {2019}
}

@ARTICLE{9448276,
  author={Z. Ding and others},
  journal={IEEE Trans. Commun.}, 
  title={No-Pain No-Gain: {DRL} Assisted Optimization in Energy-Constrained {CR-NOMA} Networks},
 month= {Sep.},
  year={2021},
  volume={69},
  number={9},
  pages={5917-5932}
}

@ARTICLE{11202497,
  author={Zhu, Guangyu and Mu, Xidong and Guo, Li and Xu, Shibiao and Liu, Yuanwei and Al-Dhahir, Naofal},
  journal={IEEE Trans. Commun.}, 
  title={Pinching-Antenna Systems ({PASS})-Enabled Secure Wireless Communications},
 month= {Oct.},
  year={2025},
  volume={74},
  number={},
  pages={490-505}}

@ARTICLE{11300296,
  author={J. Zhao and others},
  journal={IEEE Trans. Commun.}, 
  title={Pinching-Antenna Systems-Enabled Multi-User Communications: Transmission Structures and Beamforming Optimization}, 
  year={2025},
  month= {Dec.},
  volume={74},
  number={},
  pages={2316-2330}}

@ARTICLE{10981775,
  author={Ch. Ouyang and others},
  journal={IEEE Commun. Lett.}, 
  title={Array Gain for Pinching-Antenna Systems ({PASS})}, 
  year={2025},
  month= {May.},
  volume={29},
  number={6},
  pages={1471-1475}}

@inproceedings{liu2023evaluation,
  author    = {Shijie Liu},
  title     = {An Evaluation of {DDPG}, {TD3}, {SAC}, and {PPO}: Deep Reinforcement Learning Algorithms for Controlling Continuous System},
  booktitle = {Proc. Int. Conf. Data Sci., Adv. Algorithm Intell. Comput. (DAI)},
  pages     = {15--24},
  year      = {2024},
  month= {Feb.},
  publisher = {Atlantis Press}
}

@ARTICLE{11111711,
  author={Ahmadzadeh, Mohsen and others},
  journal={IEEE Internet Things J.}, 
  title={{AI}-Based Fluid Antenna Design for Client Selection in Over-the-Air Federated Learning}, 
  year={2025},
  month={Oct.},
  volume={12},
  number={20},
  pages={42549-42558}}

	\vfill
	
\end{document}